\documentclass[%
 reprint,
 superscriptaddress,
 amsmath,amssymb,
 aps,
 prc,
 twocolumn,
]{revtex4-2}

\usepackage[utf8]{inputenc} 
\usepackage{graphicx}      
\usepackage{amsmath}       
\usepackage{amssymb}       
\usepackage{hyperref}      
\usepackage{natbib}        
\usepackage{subcaption}
\usepackage[version=4]{mhchem}

\hypersetup{
    colorlinks=true,
    linkcolor=blue,
    citecolor=blue,
    urlcolor=blue
}

\begin{document}

\preprint{APS/123-QED}

\title{Static Fission Properties of Even–Even Actinides within the Warsaw Macroscopic–Microscopic Model Using Fourier-over-Spheroid Parameterization}

\author{A. Augustyn}
\email[Corresponding author: ]{aleksander.augustyn@ncbj.gov.pl}
\affiliation{National Centre for Nuclear Research, Pasteura 7, 02-093 Warsaw, Poland}

\author{T. Cap}
\affiliation{National Centre for Nuclear Research, Pasteura 7, 02-093 Warsaw, Poland}

\author{R. Capote}
\affiliation{NAPC–Nuclear Data Section, International Atomic Energy Agency, Vienna, A-1400, Austria}

\author{M. Kowal}
\affiliation{National Centre for Nuclear Research, Pasteura 7, 02-093 Warsaw, Poland}

\author{K. Pomorski}
\affiliation{National Centre for Nuclear Research, Pasteura 7, 02-093 Warsaw, Poland}

\date{\today}

\begin{abstract}
A systematic study of fission barrier heights and static properties of even–even actinide nuclei from Th to Cf has been performed within the Warsaw macroscopic–microscopic model using the five-dimensional Fourier-over-Spheroid (FoS) shape parameterization. The use of a large deformation grid, containing about $1.3\times10^{8}$ points for each nucleus, allows for a refined and numerically complete exploration of the potential-energy landscape without dividing the configuration space into subregions or applying interpolation. Barrier heights, extracted via the Immersion Water Flow method, show good agreement with empirical evaluations (including the new IAEA RIPL-4 dataset) with mean deviations below 1~MeV. Special attention is given to the long-debated third, hyperdeformed minimum. For Th isotopes, a shallow but distinct third well appears, whereas it's absent in heavier actinides (U, Pu).
\end{abstract}

\maketitle


\section{Introduction}\label{sec1}

Almost immediately after its discovery, the phenomenon of nuclear fission—whether induced by neutrons~\cite{Hahn1939} or occurring spontaneously~\cite{Flerov1940}—was interpreted through an analogy with biological cell division~\cite{Meitner1939}. In the ensuing years, the nucleus came to be viewed as a charged, incompressible liquid drop~\cite{Bohr1939a,Bohr1939}, extending the pioneering insights of Gamow on nuclear matter~\cite{Gamow1930}. This conceptual development established the foundation of the liquid-drop model, which has since become a cornerstone in understanding nuclear deformation, stability, and decay of atomic nuclei.

The concept of the \textit{fission barrier} emerged from the early liquid-drop picture of the nucleus as a charged, deformable droplet of nuclear matter, whose potential energy depends on its collective shape $\boldsymbol{q} = (q_1, q_2, \dots, q_n)$ which spans an $n$-dimensional collective coordinate space. This idea was introduced in the pioneering work of Bohr and Wheeler~\cite{Bohr1939}, who first formulated the liquid-drop model and connected nuclear deformation with the phenomenon of fission. Comprehensive expositions of nuclear collective models and macroscopic–microscopic approaches can be found in the textbooks~\cite{PomorskaPomorski2024,KrappePomorski2012} and in the recent review~\cite{KowalSkalski2022}. Within this framework, the balance between surface tension and Coulomb repulsion gives rise to a potential energy surface (PES) that governs the dynamics of nuclear deformation. This surface exhibits a local minimum corresponding to the compact ground-state configuration and rises toward a saddle region separating it from the elongated, two-lobed shapes characteristic of the pre-scission configuration. Beyond the saddle point, the potential decreases with increasing elongation as the two emerging fragments become separated. The potential energy surface $V(\boldsymbol{q})$ thus forms a multidimensional landscape connecting the ground-state minimum with the scission region. Along any admissible fission path $\mathcal{C}$ on this surface, the nucleus must overcome a potential ridge before descending into the valley leading to two separated fragments. The highest point along this path, $V_{\max}(\mathcal{C})$, defines the maximal potential energy encountered during the collective evolution. 

The fission barrier height is then defined as the minimum energy that the system must overcome when evolving from the ground state to the scission configuration,
\begin{equation}
  \label{barr1}
  B \;=\;
  \min_{\mathcal{C}}
  \Big\{
    \max_{\boldsymbol{q} \in \mathcal{C}}
    \big[\, V(\boldsymbol{q}) - E_{\text{g.s.}} \,\big]
  \Big\},
\end{equation}
where the outer minimization is taken over all continuous trajectories $\mathcal{C}$ connecting the ground-state and scission configurations in the multidimensional collective space spanned by the deformation parameters $\boldsymbol{q}$. The configuration $\boldsymbol{q}_{\text{saddle}}$ corresponding to this minimum identifies the saddle point on the potential energy surface, which serves as the gateway between the metastable nucleus and the open fission channels.

The fission barrier is one of the key quantities governing the stability of atomic nuclei. At the most fundamental level, it determines the competition between neutron evaporation and fission in compound-nucleus reactions, thereby controlling the survival probability of evaporation residues. The same barrier also dictates the systematics of spontaneous fission half-lives, as both its height and curvature determine the tunneling probability through the collective action integral~\cite{Swiatecki1955,Pomorski2022}. Consequently, it plays a decisive role in the synthesis of heavy, and especially superheavy, elements, where even a small change in $B$ by a few hundred keV can alter the predicted production cross section by several orders of magnitude. In neutron-induced or radiative-capture reactions, barrier properties control the branching between fission and other decay modes, thereby influencing the abundance patterns of heavy nuclei produced in the astrophysical $r$-process. 

Beyond its relevance to fundamental nuclear structure and nuclear reaction theory, the fission barrier is indispensable in applied nuclear physics, as it directly affects the evaluation of energy-dependent cross sections in reactor design, transmutation studies, and other practical applications. Due to this importance, fission barriers are systematically tabulated in nuclear data libraries such as RIPL-4, the latest IAEA Reference Input Parameter Library used in this paper~\cite{RIPL4}.

It must be emphasized, however, that the fission barrier height is not a universal constant. Its apparent value may depend on the excitation energy, angular momentum, or the specific reaction path under consideration—analogous to how the effective activation energy varies in complex chemical processes. This intrinsic model dependence introduces unavoidable uncertainties into theoretical descriptions, underscoring the importance of developing reliable theoretical frameworks that capture the multidimensional structure of fission barriers~\cite{Hill1953,KrappePomorski2012}. 

In this work, we present a systematic study of static fission properties for even-even actinide nuclei from Th to Cf using the Warsaw macroscopic-microscopic model with the Fourier-over-Spheroid (FoS) parametrization. This approach allows us to examine how the choice of shape representation affects fission barrier heights and hyperdeformed configurations. The calculated fission barrier heights are compared with three empirical datasets, providing a comprehensive assessment of the model's predictive capability.

%
%

\section{Macroscopic-microscopic model}\label{sec:method}

The theoretical framework employed in this study utilizes the Warsaw macroscopic-microscopic model, described in Jachimowicz \textit{et al.}~\cite{Jachimowicz2021}, which has been extensively validated for heavy and superheavy nuclei ~\cite{Jachimowicz2017_1, Jachimowicz2017_2}. In this work, we will only provide a brief summary of this method. 

Within the macroscopic-microscopic approach, the binding energy of a composite system is expressed as a sum of macroscopic and microscopic contributions:
\begin{equation}
\label{eq:binding_energy}
E_{\text{total}}(Z,N,\boldsymbol{q}) = E_{\text{macro}}(Z,N,\boldsymbol{q}) + E_{\text{micro}}(Z,N,\boldsymbol{q}),
\end{equation}
where $Z$ and $N$ denote the proton and neutron numbers, respectively, and $\boldsymbol{q}$ represents the set of nuclear shape parameters characterizing the deformation.

The macroscopic component of the binding energy  $E_{\text{macro}}(Z,N,\boldsymbol{q})$ is calculated using the Yukawa-plus-exponential model described in \cite{Krappe1979}. The microscopic energy contribution $E_{\text{micro}}(Z,N,\boldsymbol{q})$ incorporates quantum shell effects through the Strutinsky shell correction method \cite{Strutinski1967}, computed from single-particle levels of a deformed Woods-Saxon potential \cite{Cwiok1987}. Pairing correlations are treated within the Bardeen-Cooper-Schrieffer (BCS) theory. We assume a constant matrix element of the (short-range) monopole pairing interaction with values as in \cite{Muntian2001}. The pairing Hamiltonian is treated separately for neutrons and protons. It should be noted here that the macroscopic and microscopic components of the model are, to a considerable extent, adjusted independently: the former to reproduce bulk nuclear properties, and the latter to account for single-particle effects. Although this separation renders the approach somewhat inconsistent from a purely theoretical standpoint, it provides greater phenomenological flexibility than self-consistent mean-field or energy-density functional methods, in which both aspects are constrained once the effective interaction or functional is fixed (see Ref.~\cite{Jachimowicz2021} for details). In the present work, no special adjustments relative to the work~\cite{Jachimowicz2021} have been introduced to improve the agreement between the calculated fission-barrier heights and the corresponding experimental estimates.

\section{Nuclear Shape Parameterization}\label{sec:shape_param}

The selection of appropriate nuclear shape parameterization is crucial for any macroscopic–microscopic study, particularly when describing the diverse shapes adopted during the fission process. A comprehensive review of various deformation parameterizations can be found in Hasse and Myers' book~\cite{Hasse1988}.

The Warsaw macroscopic-microscopic model employs, by default, the following parametrization of the nuclear surface, where the deviation of mononuclear shapes from a sphere is expanded via spherical harmonics $Y_{\lambda}^{\mu}(\vartheta, \varphi)$:
\begin{equation}\label{eq:beta_surface}
R(\vartheta, \varphi) = c(\{\beta\}) R_0 \{ 1 + \sum_{\lambda=1}^{\infty} \sum_{\mu=-\lambda}^{+\lambda} \beta_{\lambda}^{\mu} Y_{\lambda}^{\mu} \}.
\end{equation}
Here, the expansion coefficients $\beta_{\lambda}^{\mu}$ play the role of deformation parameters, $c(\{\beta\})$ is the volume-fixing factor, and $R_0$ is the radius of a spherical nucleus. This parameterization has an inherent limitation: while numerous multipole parameters $\beta_{\lambda}^{\mu}$ are necessary to describe very elongated shapes near scission, this high dimensionality causes the deformation space to include many unphysical or superfluous shapes that play no role in the fission process. Hence, it is crucial to eliminate these non-physical parameters to minimize network size and enhance calculation speed, especially when performing systematic calculations for numerous nuclei.

To overcome these limitations, the nuclear shapes in this study are described using the Fourier-over-Spheroid (FoS) parameterization, which provides a flexible and physically intuitive framework for representing axially symmetric nuclear configurations ranging from near-spherical ground states to highly elongated fission fragments~\cite{Pomorski2023_fos}. This parameterization has proven particularly effective for describing the complex shape evolution during the fission process~\cite{Pomorski2023_fragments}.

Within the FoS framework, the nuclear surface is defined in cylindrical coordinates through the relation:
\begin{equation}\label{eq:fos_surface}
\rho^2(z, \varphi) =
\frac{R_0^2}{c}\,
f\!\left(\frac{z - z_{\text{sh}}}{z_0}\right)
\frac{1 - \eta^2}{
  1 + \eta^2 + 2\eta \cos{2\varphi}
}.
\end{equation}
where $\rho(z, \varphi)$ represents the distance from the symmetry axis to the nuclear surface at position $z$, and $R_0 = r_0 A^{1/3}$ is the radius of a sphere with an equivalent volume to the nucleus under consideration, with $r_0$ being the nuclear radius constant.

The shapes in this work are limited to axially symmetric; therefore, the non-axial parameters, the $\mu$ parameter in Eq. \ref{eq:beta_surface} and the $\eta$ parameter in Eq. \ref{eq:fos_surface}, are set to zero.

The shape function $f(u)$ is expressed as a Fourier expansion over the dimensionless coordinate $u \in [-1, 1]$:
\begin{equation}\label{eq:fos_function}
\begin{split}
f(u) = 1 - u^2 
&- \sum_{k=1}^{n}
\Big\{
  a_{2k}\cos\!\left(\tfrac{2k-1}{2}\pi u\right)
\\[-2pt]
&\quad
 + a_{2k+1}\sin(k\pi u)
\Big\},
\end{split}
\end{equation}
where the first two terms describe a spheroid, while the expansion coefficients $a_i$  delineate the deviation of the nuclear surface from the spheroid and serve as deformation parameters characterizing the nuclear shape. The elongation parameter $c$ determines the half-length of the nucleus through $z_0 = cR_0$. The parameter $a_3$ describes reflection-asymmetric deformations, which become important near the scission configuration where the nascent fragments may have different masses. The parameter $a_4$ controls the waist of the nucleus and the development of its neck at larger elongations, a critical feature in the fission process. Higher-order terms, $a_5$ and $a_6$, provide additional flexibility in describing fine structural details of the nuclear surface.

Volume conservation imposes a constraint on the $a_2$ coefficient:
\begin{equation}\label{eq:volume_constraint}
a_2 = \sum_{n=2}^{\infty} \frac{(-1)^n a_{2n}}{2n-1},
\end{equation}
which reduces the number of independent parameters. The center of mass is maintained at the coordinate origin through an appropriate shift $z_{\text{sh}}$, given by:
\begin{equation}\label{eq:com_shift}
z_{\text{sh}} = \frac{3}{2\pi}z_0 \sum_{n=1}^{\infty} \frac{(-1)^{n+1} a_{2n+1}}{n}.
\end{equation}
The five-parameter space $(c, a_3, a_4, a_5, a_6)$ in the present work enables a more refined description of nuclear shapes, particularly in regions where subtle deformation effects play a significant role. This expanded parameter space captures higher-order shape variations that may influence barrier heights, providing a more comprehensive representation of the nuclear energy landscape. The computational cost of this enhancement is substantial, as the number of calculated configurations scales with the fifth power of the grid resolution, but the resulting improvement in accuracy justifies this investment for systematic studies of fission properties.

For the present calculations, we constructed five-dimensional potential energy surfaces spanning the parameter space:
\begin{center}
\begin{tabular}{lcl}
$c$   & = & $+1.00~(0.01)~2.00$ \\
$a_3$ & = & $+0.00~(0.01)~0.25$ \\
$a_4$ & = & $-0.15~(0.01)~0.35$ \\
$a_5$ & = & $-0.15~(0.01)~0.15$ \\
$a_6$ & = & $-0.15~(0.01)~0.15$ \\
\end{tabular}
\end{center}
where the values in parentheses indicate the grid spacing. This corresponds to $101\times26\times51\times31\times31 = 128,702,886$ nuclear shape configurations for each nucleus, representing a substantial computational undertaking that was executed on the Świerk Computing Centre high-performance computing infrastructure. The deformation grid used in the present calculations is exceptionally dense, ensuring that no interpolation is required between neighboring mesh points. This eliminates any numerical uncertainty associated with interpolation procedures and guarantees that all minima and saddle points are identified directly from the computed data.

\begin{figure*}[htbp]
  \centering
  \begin{subfigure}[b]{0.49\textwidth}
    \centering
    \includegraphics[width=\textwidth]{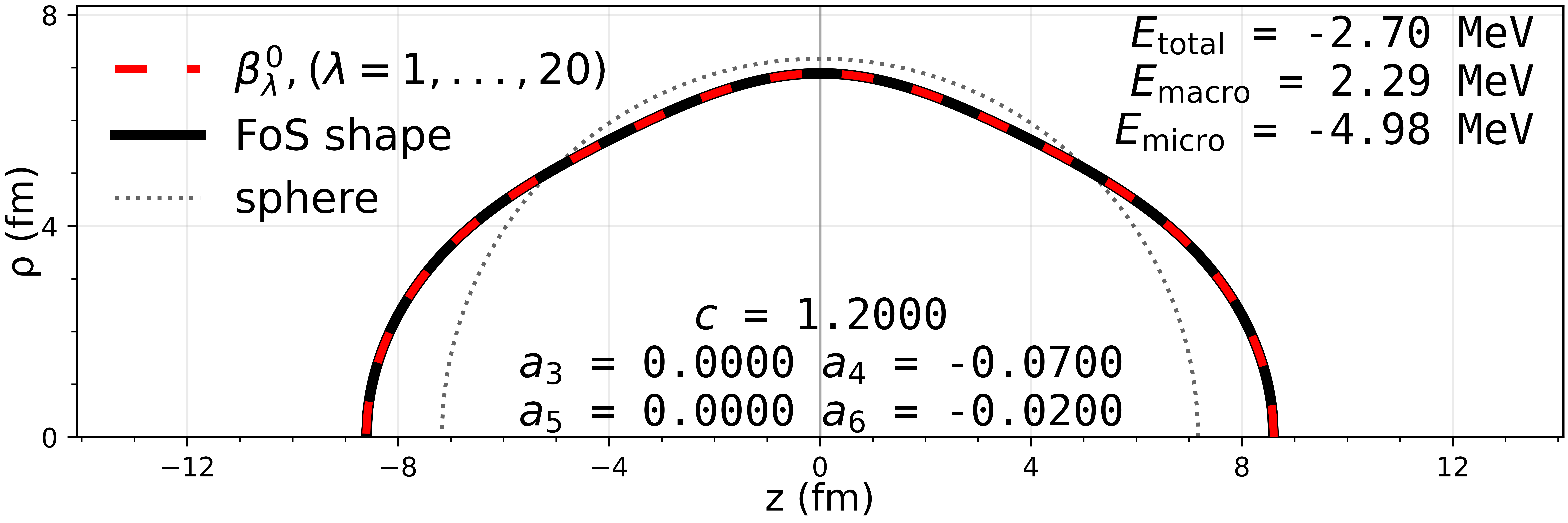}
    \caption{Ground state}
    \label{fig:u236:gs}
  \end{subfigure}
  \hfill
  \begin{subfigure}[b]{0.49\textwidth}
    \centering
    \includegraphics[width=\textwidth]{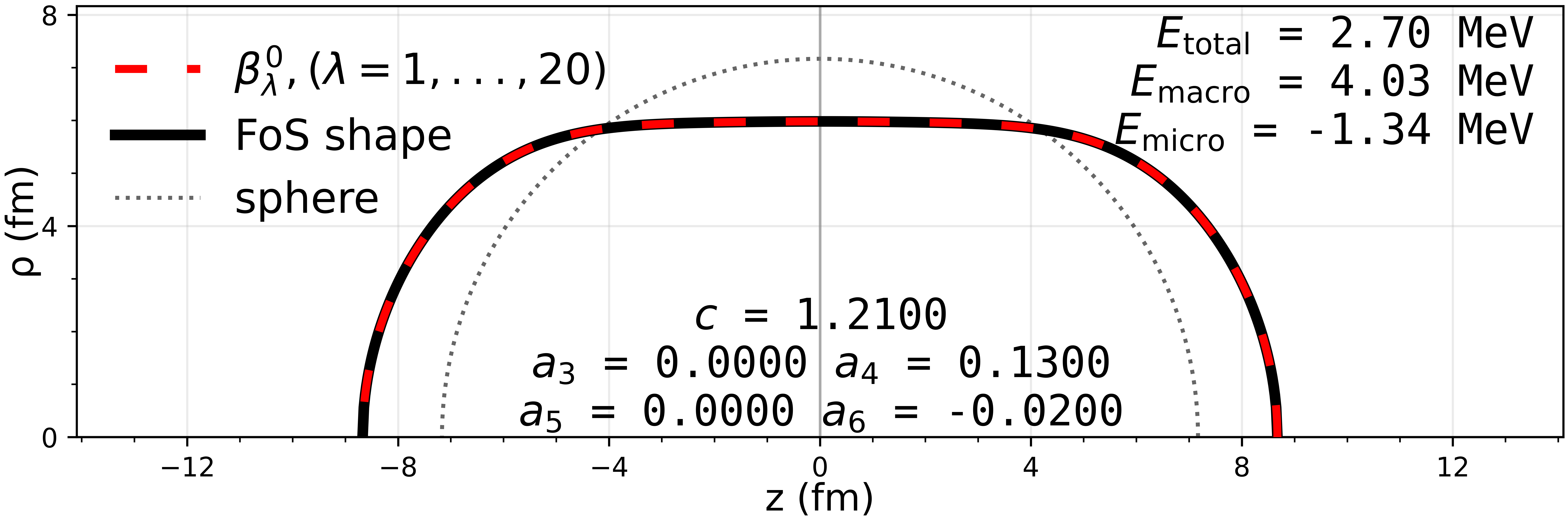}
    \caption{Inner saddle point}
    \label{fig:u236:saddle1}
  \end{subfigure}
  
  \begin{subfigure}[b]{0.49\textwidth}
    \centering
    \includegraphics[width=\textwidth]{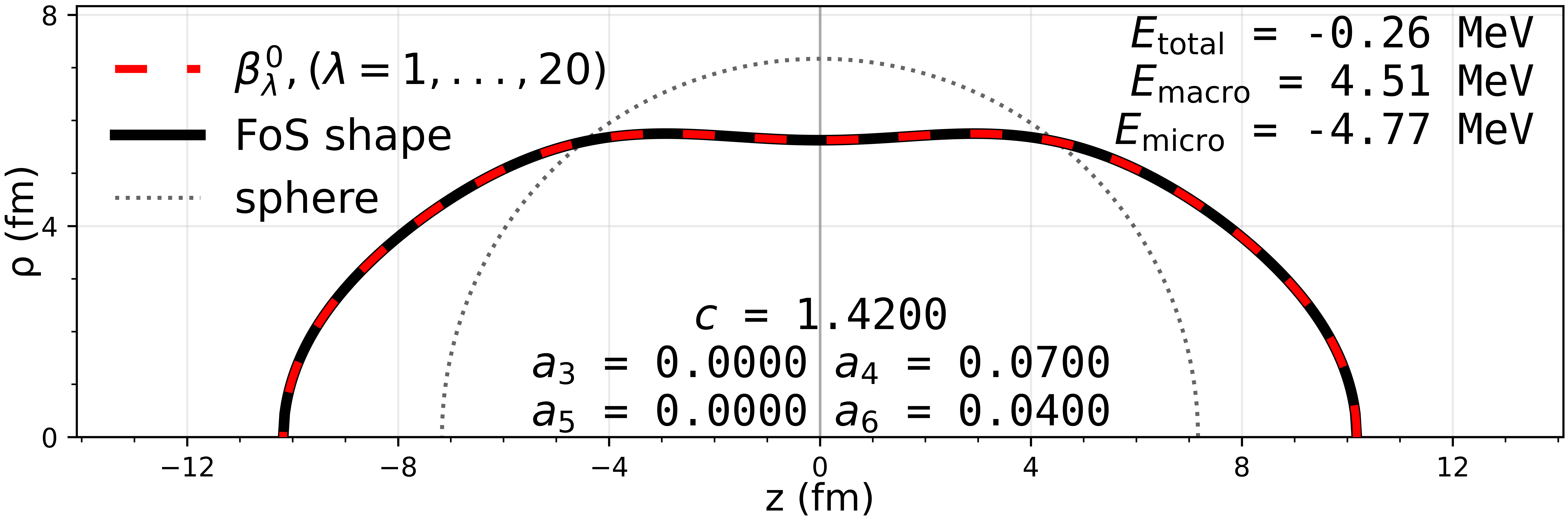}
    \caption{Secondary minimum}
    \label{fig:u236:secmin}
  \end{subfigure}
  \hfill
  \begin{subfigure}[b]{0.49\textwidth}
    \centering
    \includegraphics[width=\textwidth]{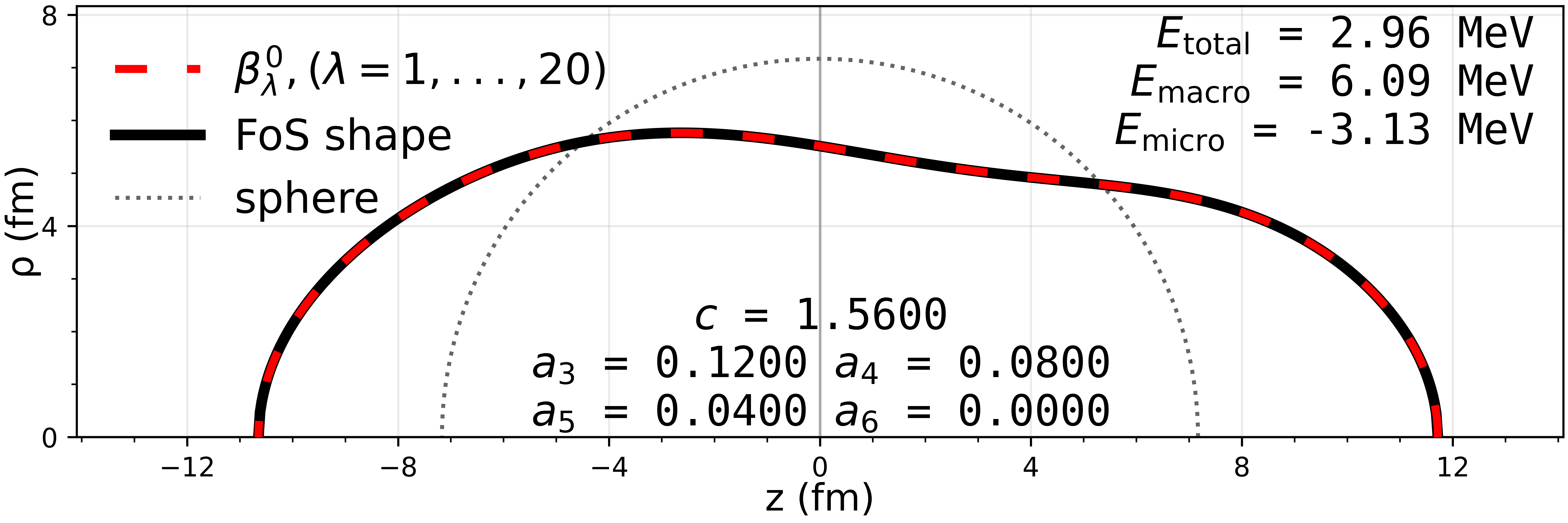}
    \caption{Outer saddle point}
    \label{fig:u236:saddle2}
  \end{subfigure}
  
  \caption{Comparison of FoS shapes (black lines) with their equivalent representation in $\beta$ parameterization (red dashed lines) for selected equilibrium points (provided below each panel) for \ce{^{236}U}. The values of parameters $c$, $a_{3}$, $a_{4}$, $a_{5}$, and $a_{6}$ are given at each point, along with the values of $E_\text{total}$, $E_\text{macro}$, and $E_\text{micro}$, representing the total, macroscopic, and microscopic energy, respectively.}
  \label{fig:shape_plots}
\end{figure*}

The numerical code of the Warsaw macroscopic–microscopic model has been extensively optimized to work with the parameterization described by Eq. \ref{eq:beta_surface}, regarding the minimization process and grid calculations. To utilize the model with FoS parameterization, the shapes must be expressed in terms of the equivalent deformation parameters from the multipole expansion. Given a radius vector $\tilde{R}(\vartheta, \varphi)$ representing a specific shape in any parameterization, including FoS, the corresponding $\beta_{\lambda}^{\mu}$ parameters can be determined using the orthogonality of spherical harmonics:
\begin{equation}\label{eq:harm_orth}
\beta_{\lambda}^{\mu} = \sqrt{4 \pi} \frac{\smallint{\tilde{R}(\vartheta, \varphi) Y_{\lambda}^{\mu} \sin\vartheta \text{d}\vartheta \text{d}\varphi}}{\smallint{\tilde{R}(\vartheta, \varphi) Y_{0}^{0} \sin\vartheta \text{d}\vartheta \text{d}\varphi}}.
\end{equation}
This methodology has been demonstrated in \cite{Moller2009}, and also used in \cite{Jachimowicz2024}. In this work, the values obtained from Eq. \ref{eq:harm_orth} are used as initial guesses and additionally optimized using analytical gradients to minimize the root mean square error. Remarkably, the transformation between the FoS parameterization and the multipole expansion remains highly effective even for extensively elongated fragments. It is worth mentioning that our updated numerical code can now consider up to $\lambda = 20$ multipoles in the case of axial symmetry, ensuring nearly exact integration and excellent reproduction of FoS shapes (see Fig.~\ref{fig:shape_plots}). Higher-order multipole deformations typically contribute very little to the total energy. Importantly, this enhancement does not introduce any additional computational overhead since the dimensionality of the deformation space is determined solely by the five FoS parameters.

By utilizing up to twenty $\beta^{0}$ parameters, we obtain practically indistinguishable shapes. It is essential to note that the current analysis does not account for triaxiality, which has significant effects in the vicinity of the inner barrier for the investigated actinide nuclei. Incorporating this parameter is a more intricate task and will be the focus of future investigations.

\section{Results}\label{sec:results}

\begin{figure*}[htbp]
  \centering
  \begin{subfigure}[b]{0.49\textwidth}
    \centering
    \includegraphics[width=\textwidth]{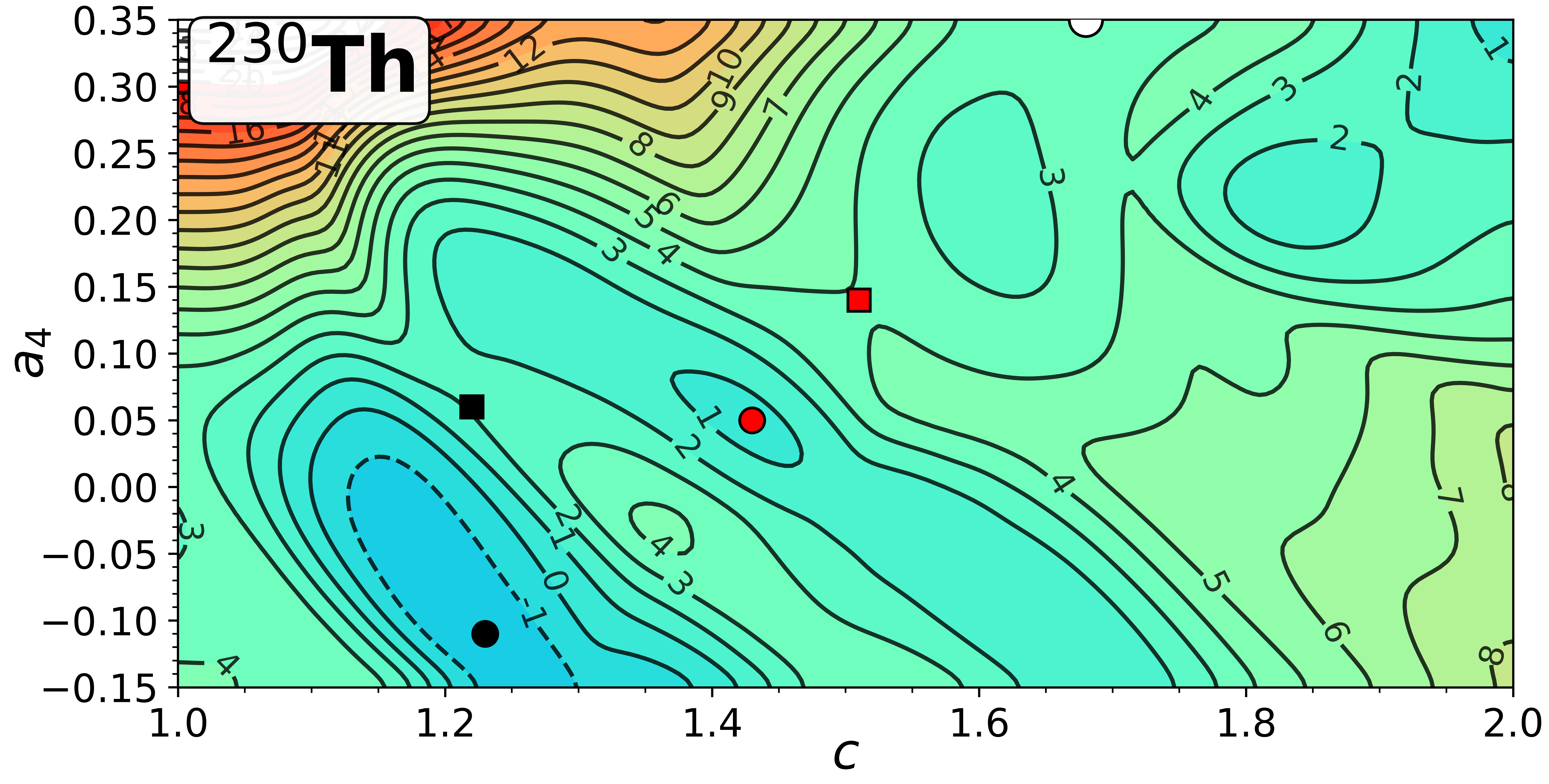}
    \label{fig:th230:map}
  \end{subfigure}
  \hfill
  \begin{subfigure}[b]{0.49\textwidth}
    \centering
    \includegraphics[width=\textwidth]{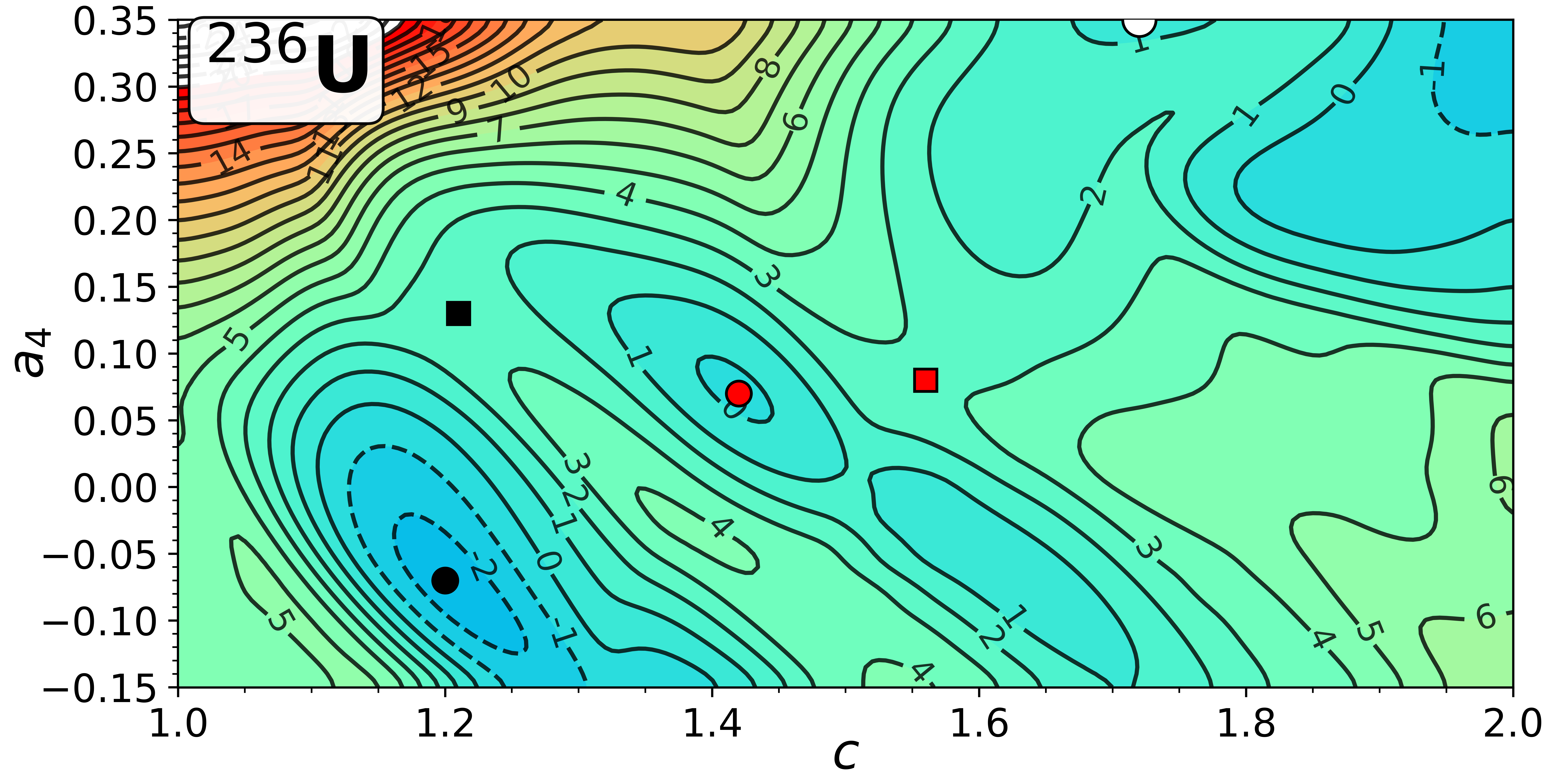}
    \label{fig:u236:map}
  \end{subfigure}
  
  \begin{subfigure}[b]{0.49\textwidth}
    \centering
    \includegraphics[width=\textwidth]{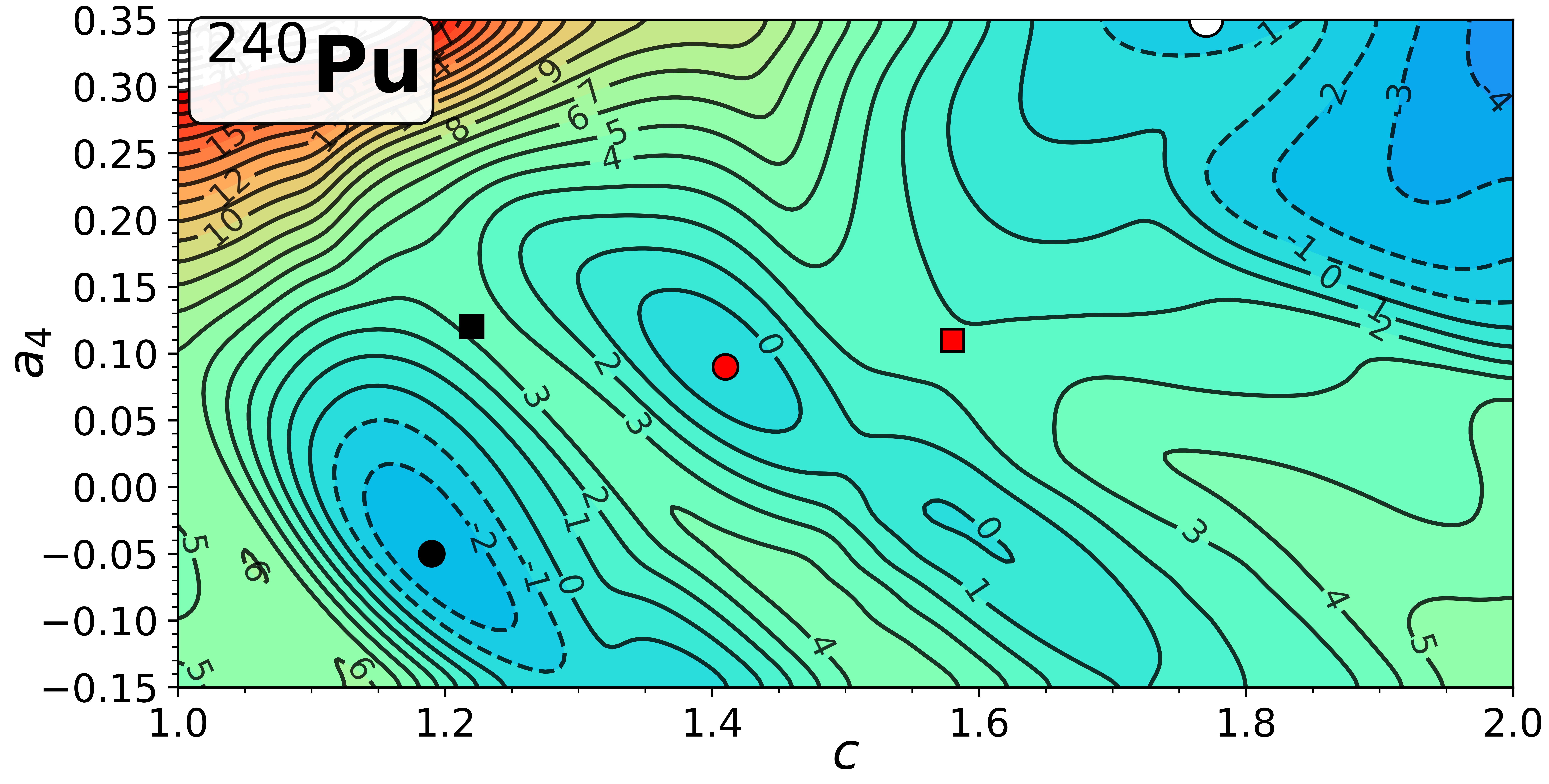}
    \label{fig:pu240:map}
  \end{subfigure}
  \hfill
  \begin{subfigure}[b]{0.49\textwidth}
    \centering
    \includegraphics[width=\textwidth]{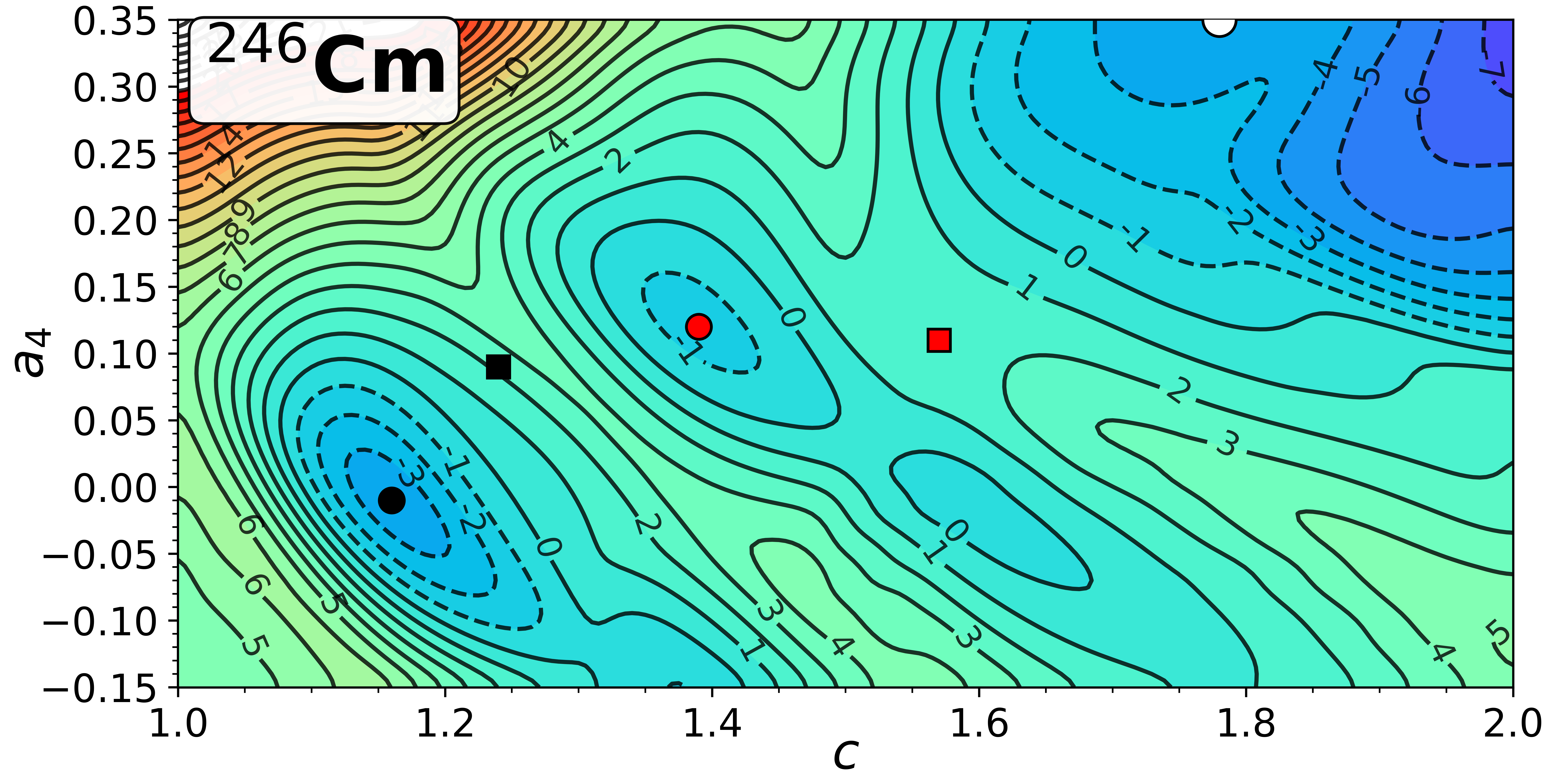}
    \label{fig:cm246:map}
  \end{subfigure}
  
  \caption{Potential energy surfaces in the $\{c, a_4\}$ plane for \ce{^{230}Th}, \ce{^{236}U}, \ce{^{240}Pu}, and \ce{^{246}Cm} nuclei, obtained by minimization over the remaining parameters: $a_3$, $a_5$, and $a_6$. Total energy $E_{\text{total}}$ (in MeV) is given relative to the macroscopic energy of the spherical configuration.}
  \label{fig:maps}
\end{figure*}

\begin{figure*}[htbp]
  \centering
  \includegraphics[width=\textwidth]{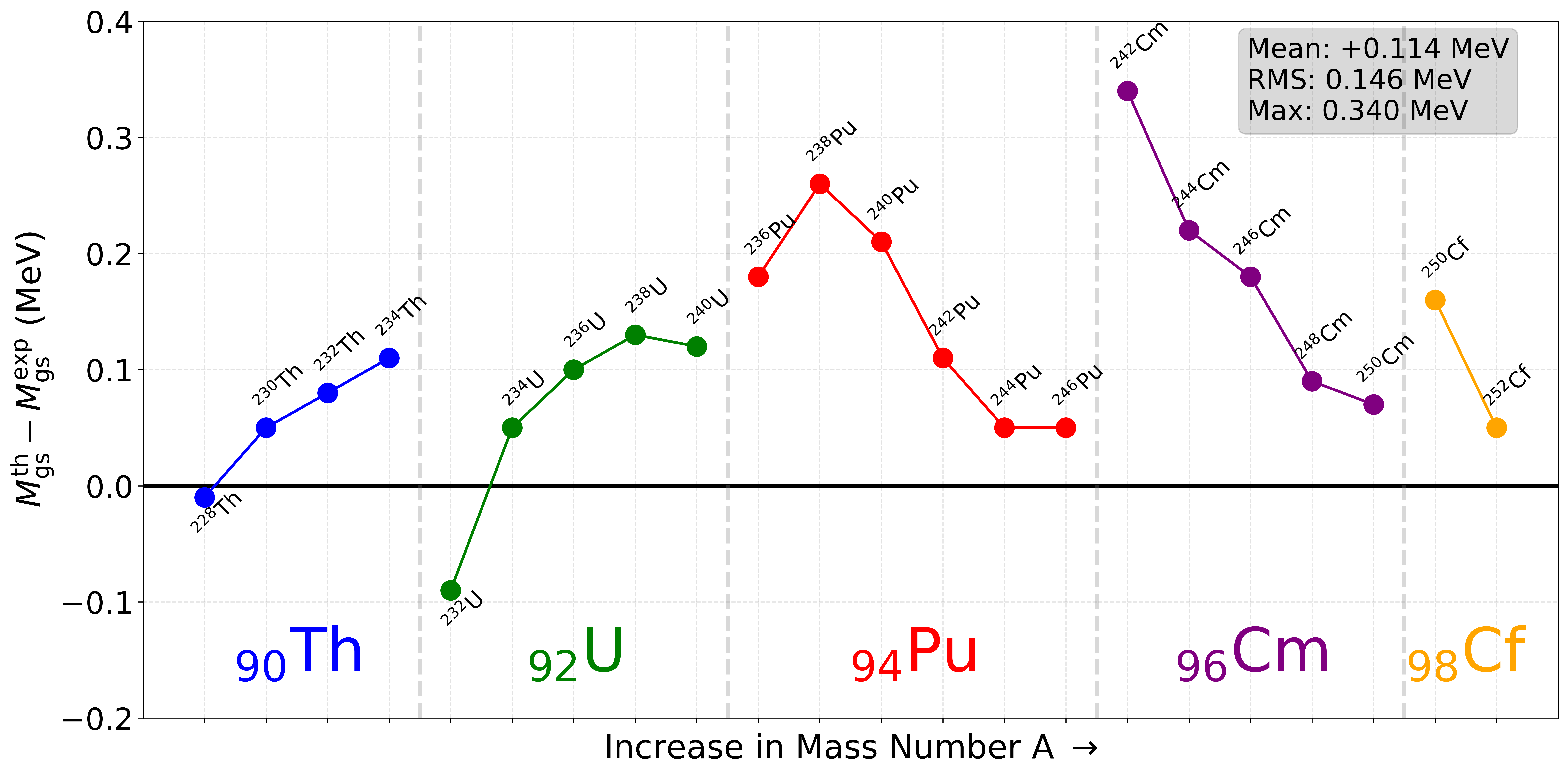}
  \caption{Differences between calculated and experimental ground-state mass excesses for even–even actinides, $\Delta M = M^\text{th}_\text{gs} - M^\text{exp}_\text{gs}$. Theoretical mass excesses are obtained within the Warsaw macroscopic–microscopic model employing the five-dimensional Fourier-over-Spheroid (FoS) parameterization. Empirical reference data are taken from the AME2020 mass evaluation~\cite{Audi2020}.}
\label{fig:mass_excess}
\end{figure*}

\begin{figure*}[htbp]
  \centering
  \includegraphics[width=\textwidth]{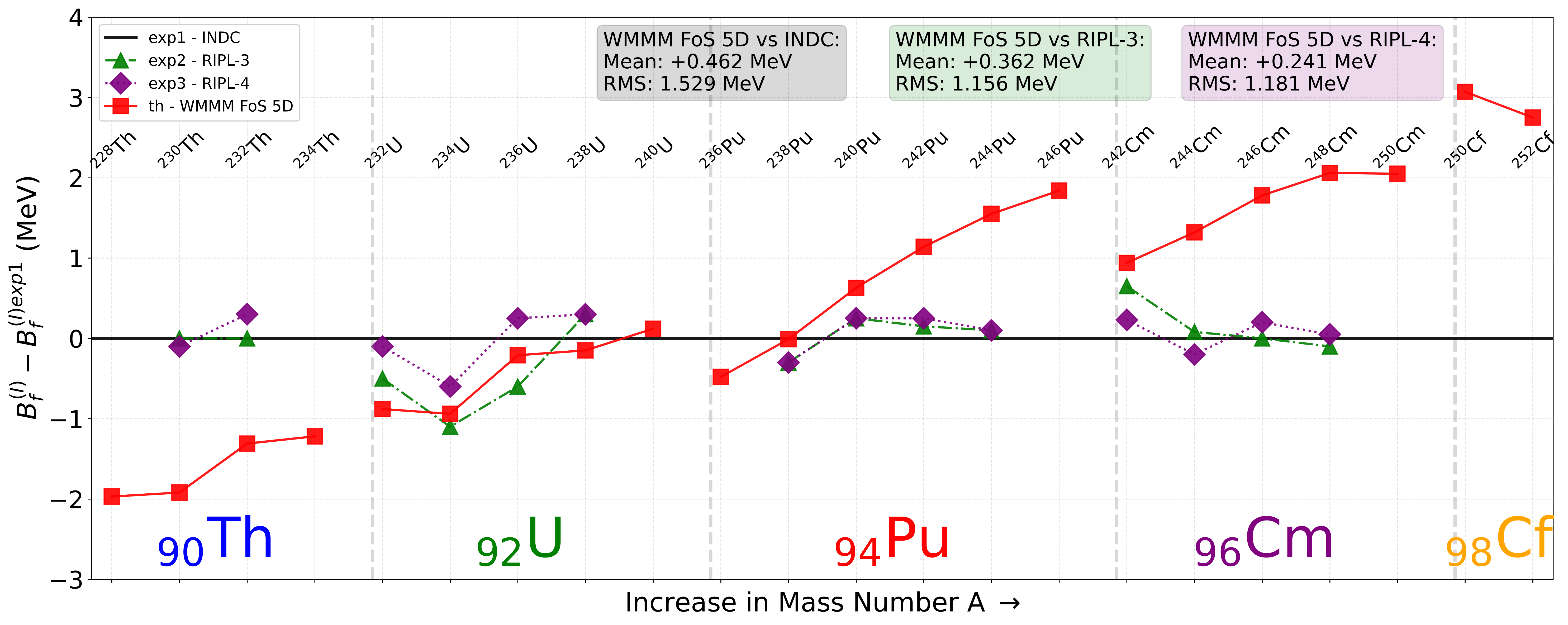}
  \caption{The difference $B_{f}^{\rm{(I)}} - B_{f}^{\rm{(I)exp1}}$, allowing direct comparison with the INDC~\cite{Smirenkin1993} evaluation ($B_{f}^{\rm{(I)exp1}}$) as a reference. Here, $B_{f}^{\rm{(I)}}$ is either the theoretical inner barrier height ($B_{f}^{\rm{(I)th}}$) from this work or the empirical value from RIPL-3 ($B_{f}^{\rm{(I)exp2}}$)~\cite{Capote2009} or RIPL-4 ($B_{f}^{\rm{(I)exp3}}$)~\cite{RIPL4}. The INDC dataset is chosen as the reference point, as it contains all analyzed systems. See Tab.~\ref{tab:results} for details. The theoretical values were obtained within the Warsaw macroscopic–microscopic model using the five-dimensional FoS parameterization (WMMM FoS 5D) without non-axial effects.
 }
  \label{fig:inner_barrier}
\end{figure*}

\begin{figure*}[htbp]
  \centering
  \includegraphics[width=\textwidth]{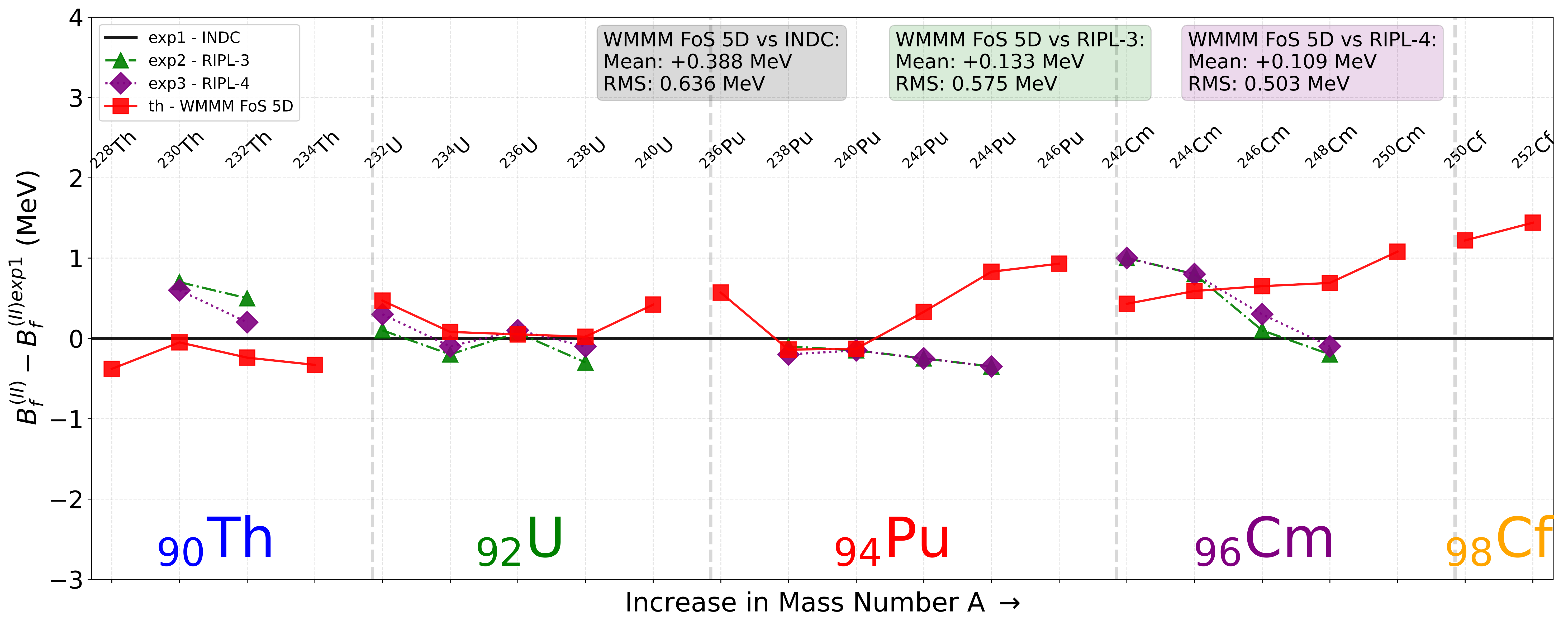}
  \caption{Same as in Fig.~\ref{fig:inner_barrier}, but for the outer fission barrier $B_{f}^{\rm{(II)}}$ heights for even–even actinide nuclei.}
  \label{fig:outer_barrier}
\end{figure*}

\begin{table*}[htbp]
\centering
\caption{
Comparison of ground-state mass excesses: calculated $M^\text{th}_\text{gs}$ and measured $M^\text{exp}_\text{gs}$ from the AME2020 evaluation~\cite{Audi2020}. Calculated inner $B^\text{(I)th}_\text{f}$ and outer $B^\text{(II)th}_\text{f}$ fission barrier heights, and three corresponding empirical data sets: (I)~$B^\text{(I)exp1}_\text{f}$, $B^\text{(II)exp1}_\text{f}$ from Smirenkin \textit{et al.}~\cite{Smirenkin1993}, (II)~$B^\text{(I)exp2}_\text{f}$, $B^\text{(II)exp2}_\text{f}$ recommended in the IAEA RIPL-3 project~\cite{Capote2009}, and (III)~$B^\text{(I)exp3}_\text{f}$, $B^\text{(II)exp3}_\text{f}$ recommended in the IAEA RIPL-4 project~\cite{RIPL4}. All quantities are in MeV.}
\label{tab:results}
\begin{tabular}{lllllllllll}
\hline
Nucleus            & $M^\text{th}_\text{gs}$ & $M^\text{exp}_\text{gs}$ & $B^\text{(I)th}_\text{f}$ & $B^\text{(I)exp1}_\text{f}$ & $B^\text{(I)exp2}_\text{f}$ & $B^\text{(I)exp3}_\text{f}$ & $B^\text{(II)th}_\text{f}$ & $B^\text{(II)exp1}_\text{f}$ & $B^\text{(II)exp2}_\text{f}$ & $B^\text{(II)exp3}_\text{f}$\\
\hline \hline
\ce{^{228}Th} & 26.76               & 26.77          & 4.23                  & 6.20              & \textbf{{ }---{ }}               & \textbf{{ }---{ }}               & 6.12                   & 6.50               & \textbf{{ }---{ }}                & \textbf{{ }---{ }}                \\
\ce{^{230}Th} & 30.91               & 30.86          & 4.18                  & 6.10              & 6.10              & 6.00              & 6.05                   & 6.10               & 6.80               & 6.70               \\
\ce{^{232}Th} & 35.53               & 35.45          & 4.49                  & 5.80              & 5.80              & 6.10              & 5.96                   & 6.20               & 6.70               & 6.40               \\
\ce{^{234}Th} & 40.72               & 40.61          & 4.88                  & 6.10              & \textbf{{ }---{ }}               & \textbf{{ }---{ }}               & 5.97                   & 6.30               & \textbf{{ }---{ }}                & \textbf{{ }---{ }}                \\
\ce{^{232}U}  & 34.52               & 34.61          & 4.52                  & 5.40              & 4.90              & 5.30              & 5.77                   & 5.30               & 5.40               & 5.60               \\
\ce{^{234}U}  & 38.20               & 38.15          & 4.96                  & 5.90              & 4.80              & 5.30              & 5.78                   & 5.70               & 5.50               & 5.60               \\
\ce{^{236}U}  & 42.55               & 42.45          & 5.39                  & 5.60              & 5.00              & 5.85              & 5.65                   & 5.60               & 5.67               & 5.70               \\
\ce{^{238}U}  & 47.44               & 47.31          & 5.85                  & 6.00              & 6.30              & 6.30              & 5.82                   & 5.80               & 5.50               & 5.70               \\
\ce{^{240}U}  & 52.84               & 52.72          & 6.22                  & 6.10              & \textbf{{ }---{ }}               & \textbf{{ }---{ }}               & 6.22                   & 5.80               & \textbf{{ }---{ }}                & \textbf{{ }---{ }}                \\
\ce{^{236}Pu} & 43.08               & 42.90          & 5.22                  & 5.70              & \textbf{{ }---{ }}               & \textbf{{ }---{ }}               & 5.07                   & 4.50               & \textbf{{ }---{ }}                & \textbf{{ }---{ }}                \\
\ce{^{238}Pu} & 46.42               & 46.17          & 5.89                  & 5.90              & 5.60              & 5.60              & 5.06                   & 5.20               & 5.10               & 5.00               \\
\ce{^{240}Pu} & 50.34               & 50.13          & 6.43                  & 5.80              & 6.05              & 6.05              & 5.17                   & 5.30               & 5.15               & 5.15               \\
\ce{^{242}Pu} & 54.83               & 54.72          & 6.84                  & 5.70              & 5.85              & 5.95              & 5.63                   & 5.30               & 5.05               & 5.05               \\
\ce{^{244}Pu} & 59.86               & 59.81          & 7.15                  & 5.60              & 5.70              & 5.70              & 6.03                   & 5.20               & 4.85               & 4.85               \\
\ce{^{246}Pu} & 65.45               & 65.40          & 7.24                  & 5.40              & \textbf{{ }---{ }}               & \textbf{{ }---{ }}               & 6.23                   & 5.30               & \textbf{{ }---{ }}                & \textbf{{ }---{ }}                \\
\ce{^{242}Cm} & 55.14               & 54.81          & 6.94                  & 6.00              & 6.65              & 6.23              & 4.43                   & 4.00               & 5.00               & 5.00               \\
\ce{^{244}Cm} & 58.67               & 58.45          & 7.42                  & 6.10              & 6.18              & 5.90              & 4.89                   & 4.30               & 5.10               & 5.10               \\
\ce{^{246}Cm} & 62.80               & 62.62          & 7.78                  & 6.00              & 6.00              & 6.20              & 5.35                   & 4.70               & 4.80               & 5.00               \\
\ce{^{248}Cm} & 67.48               & 67.39          & 7.96                  & 5.90              & 5.80              & 5.95              & 5.69                   & 5.00               & 4.80               & 4.90               \\
\ce{^{250}Cm} & 73.06               & 72.99          & 7.45                  & 5.40              & \textbf{{ }---{ }}               & \textbf{{ }---{ }}               & 5.48                   & 4.40               & \textbf{{ }---{ }}                & \textbf{{ }---{ }}                \\
\ce{^{250}Cf} & 71.33               & 71.17          & 8.67                  & 5.60              & \textbf{{ }---{ }}               & \textbf{{ }---{ }}               & 5.02                   & 3.80               & \textbf{{ }---{ }}                & \textbf{{ }---{ }}                \\
\ce{^{252}Cf} & 76.09               & 76.03          & 8.05                  & 5.30              & \textbf{{ }---{ }}               & \textbf{{ }---{ }}               & 4.94                   & 3.50               & \textbf{{ }---{ }}                & \textbf{{ }---{ }}             \\ \hline
\end{tabular}
\end{table*}

Figure~\ref{fig:maps} shows the total potential energy surfaces for 4 selected nuclei calculated using the FoS shape parametrization. The maps are derived by projecting the full 5D potential energy surface onto two selected coordinates (in this case, elongation $c$ and neck variable $a_4$), while minimizing the energy over the remaining degrees of freedom. Minima are marked as circles: ground state (black), secondary minimum (red), fission exit (white). The fission exit is a local minimum on the $a_4$ boundary with the lowest value of $c$, and it denotes a point through which the system escapes the PES. Saddle points are marked as squares: inner saddle (black), outer saddle (red).

The inner and outer fission barriers were identified using a modified version of the Immersion Water Flow (IWF) algorithm. Since the complete five-dimensional potential-energy surface (PES) was precomputed on a dense grid without interpolation, the procedure operates directly on the discrete deformation mesh. In our implementation: (i) two minima between which we seek the saddle point are selected, (ii) all grid points are sorted in ascending order of energy, (iii) points are processed sequentially, with each activated point merged into connected basins using a disjoint-set (union-find) data structure, (iv) a saddle point is identified when, upon processing, the basins containing the two selected minima first become connected, and (v) the barrier height is defined as the energy difference between the saddle and the ground state. In the case of the inner fission barrier, the two minima between which the saddle point is searched for are the ground state and the second minimum. In the case of the outer fission barrier, the secondary minimum and the fission exit point are selected.

Because the PES is fully discrete, the method does not require gradients or interpolation and is therefore numerically robust. Neighborhood relations include all nearest orthogonal grid points along the five deformation coordinates (10 neighbors per point in the five-dimensional space), ensuring correct topological connectivity in the high-dimensional landscape. This procedure yields well-defined inner and outer saddle points that separate the ground-state minimum, the superdeformed (second) minimum, and the descent toward scission.

The maps in Fig.~\ref{fig:maps} clearly depict the ground states and secondary minima, while the barriers separating these minima are visualized through the contour topography. All four nuclei exhibit well-defined ground state minima (black circles) located at moderate elongation ($c \approx 1.2-1.25$) with small negative $a_4$ values, residing in deep potential wells indicated by the dense contour lines. The secondary minima (red circles) appear at higher elongations ($c \approx 1.35-1.45$) and show systematic variation across the isotopic chain. Notably, $^{246}$Cm displays a particularly pronounced secondary minimum with comparable depth to its ground state, while the lighter nuclei show progressively shallower secondary wells.

The fission barriers are clearly visible as ridges in the potential energy landscape. The inner saddle points (black squares) define the inner barriers at $c \approx 1.2$ and moderate positive $a_4$ values, while the second saddle points (red squares) at $c \approx 1.5-1.6$ mark the outer barriers. The barrier heights can be inferred from the contour spacing, with $^{246}$Cm exhibiting the most complex barrier structure, including a prominent ridge that extends to large positive $a_4$ values. The transition from prolate to superdeformed configurations is evident in the energy gradient between the two minima, with the barrier topography becoming increasingly asymmetric for heavier actinides.

We note that the saddle points marked in Fig. \ref{fig:maps} correspond to the true stationary configurations identified on the full five-dimensional potential-energy surface (PES) by means of the Immersion Water Flow (IWF) algorithm. In contrast, the two-dimensional maps shown in the figure represent projected energy surfaces obtained after minimization over the remaining deformation parameters. Because a projection with constrained minimization does not preserve the locations of stationary points, the saddles visible on the two-dimensional maps do not, in general, coincide with the genuine five-dimensional saddle configurations. This slight displacement is a natural geometric consequence of reducing a multidimensional surface to a lower-dimensional representation: a saddle point of the full PES need not appear as a saddle in any two-dimensional section or projection.

\subsection{Empirical barriers derived from measured fission cross sections}
A direct comparison between theoretically calculated and empirically inferred fission barrier heights constitutes an intrinsically delicate task. Theoretical treatments, in particular those employing the potential energy surface (PES) framework, probe an inherently multidimensional collective space characterized by a rich manifold of shape degrees of freedom, encompassing elongation, mass asymmetry, neck curvature, and higher-order multipole deformations. By contrast, empirical extractions of barrier parameters from reaction calculations are most commonly based on phenomenological nuclear reaction analyses within effectively one-dimensional tunneling schemes, which, by construction, neglect the interdependence and coupling among these collective coordinates. This fundamental disparity in dimensionality between the multidimensional topography of theoretical potential landscapes and the reduced-dimensional projections accessible to experiment inevitably engenders systematic ambiguities and limits the precision of any direct confrontation between these theoretical estimates and the empirical estimates derived from measured fission cross sections. 

However, the measured fission cross sections may exhibit fission threshold behavior (e.g., for even-$A$ U isotopes), which can be directly linked to the height of the highest fission barrier and represents the only model-independent indication of the actual barrier height. Resonances in the measured fission cross section may also provide information about the barrier structure, and the depth of the super- and hyper-deformed deformation surface minima. Therefore, comparing empirical fission barriers derived from reaction studies with theoretical ones often reveals complementary information, making it a challenging task for any theoretical study.  

For comparison with our calculations, we selected three empirical sets of fission barriers derived from fission cross section measurements: one proposed by Smirenkin and collaborators \cite{Smirenkin1993}, and the two sets recommended in IAEA RIPL-3 \cite{Capote2009} and RIPL-4 projects \cite{RIPL4}. The RIPL-3 recommendations \cite{Capote2009} have been widely used to compare with theoretical fission barriers and supersede Smirenkin barriers \cite{Smirenkin1993} for actinide nuclei. By comparing the three different empirical barrier sets, we can gain insight into the typical uncertainties associated with extracting barriers from measured fission cross section data, as reflected in the spread of those empirical barriers. 

RIPL-4 recommendations \cite{RIPL4} are still unpublished, so a brief description of how they were derived is relevant. The group responsible for the RIPL dataset reviewed published available information on estimated fission barrier properties, including compilations by Bj{\o}rnholm and Lynn \cite{Bjornholm80}, Smirenkin \cite{Smirenkin1993}, and Maslov \cite{Maslov98}, and the fission barriers used in the cross section calculations to produce the JENDL-5 evaluated library \cite{Iwamoto23}. The RIPL-4 review aimed to define double-humped empirical barriers, which can serve as a starting point for modeling the measured fission cross sections. The RIPL-4 recommendations \cite{RIPL4} were validated by performing fission cross section calculations based on those empirical barriers using the EMPIRE modeling code system \cite{Herman07}, following the same approach as the model calculations used for nuclear data evaluation (e.g., \cite{Capote18}). 

\subsection{Comparison with empirical barriers}

Analysis was performed for all 22 even-even actinide nuclei of Th, U, Pu, Cm, and Cf included in the INDC dataset~\cite{Smirenkin1993}. For completeness and potential use in future applications, all numerical values extracted from the calculated potential-energy maps are summarized in Table~\ref{tab:results}. The table lists the calculated ground-state mass excesses $M^\text{th}_\text{gs}$, as well as the inner $B^\text{(I)}_\text{f}$ and outer fission barrier heights $B^\text{(II)}_\text{f}$ for each even--even actinide nucleus considered in this work. The calculations are compared with experimental mass excesses $M^\text{exp}_\text{gs}$ from the AME2020 atomic mass evaluation~\cite{Audi2020} and the empirical barrier heights for all three selected data sets described in the previous section~\cite{Smirenkin1993,Capote2009,RIPL4}.

The calculated ground state mass excesses are displayed in Fig.~\ref{fig:mass_excess}, and demonstrate excellent agreement with experimental values from the atomic mass evaluation~\cite{Audi2020}. The mean deviation of $+0.114$ MeV with a root-mean-square error (RMSE) of $0.146$ MeV indicates that the Warsaw macroscopic-microscopic model with the five-dimensional FoS parameterization successfully reproduces the binding energies of these systems. This level of agreement provides confidence that the model captures the essential physics governing nuclear stability in the actinide region.

We now turn to the comparison between the different types of barriers. The inner barrier corresponds to the more compact, less deformed configuration separating the ground state from the superdeformed outer minimum. The outer barrier, by contrast, is associated with much larger elongations and thinner necks, and it separates the superdeformed minimum (or third minimum) from the scission region where the nucleus eventually splits into two fragments. It should be emphasized that the terms “inner” and “outer” refer to the degree of nuclear deformation rather than to the order of appearance in the potential profile. They are therefore not strictly synonymous with “primary” and “secondary” barriers — the latter terminology denotes, respectively, the higher and lower barriers in energy, irrespective of the underlying shape deformation. This distinction is important when comparing theoretical results with empirical extractions, where the labeling conventions may differ depending on the adopted fitting procedure.

The inner fission barrier heights are presented in Fig.~\ref{fig:inner_barrier} and show reasonable agreement with empirical data from the INDC compilation~\cite{Smirenkin1993} and the RIPL-3~\cite{Capote2009} and RIPL-4 evaluations \cite{RIPL4}. 
The calculated inner barriers exhibit smooth isotopic trends from Th to Cf and exhibit a mean positive offset of $+0.462$ MeV relative to the INDC data, with a root-mean-square deviation of $1.529$ MeV. The differences most likely stem from the current approach not accounting for non-axial deformations. It is worth noting that the inner fission barrier height is often poorly determined in empirical estimates and typically correspond to the "secondary" barrier.  

The outer fission barrier heights, shown in Fig.~\ref{fig:outer_barrier}, demonstrate closer agreement with empirical data than the inner barriers, with a mean deviation of $+0.388$ MeV and a root-mean-square error of $0.636$ MeV when compared to the INDC compilation. The calculated outer barriers gradually decrease in height with increasing neutron number. This improved agreement suggests a reduced effect of nonaxiality in this region. However, the same comment as above applies, the uncertainty of the empirically determined "primary" (highest barrier) is much lower than the one for the corresponding inner barrier. Therefore, the better agreement with empirical data for outer barriers is expected and provides a more stringent test of the theoretical model. 

\begin{figure*}[htbp]
  \centering
  \begin{subfigure}[b]{0.49\textwidth}
    \centering
    \includegraphics[width=\textwidth]{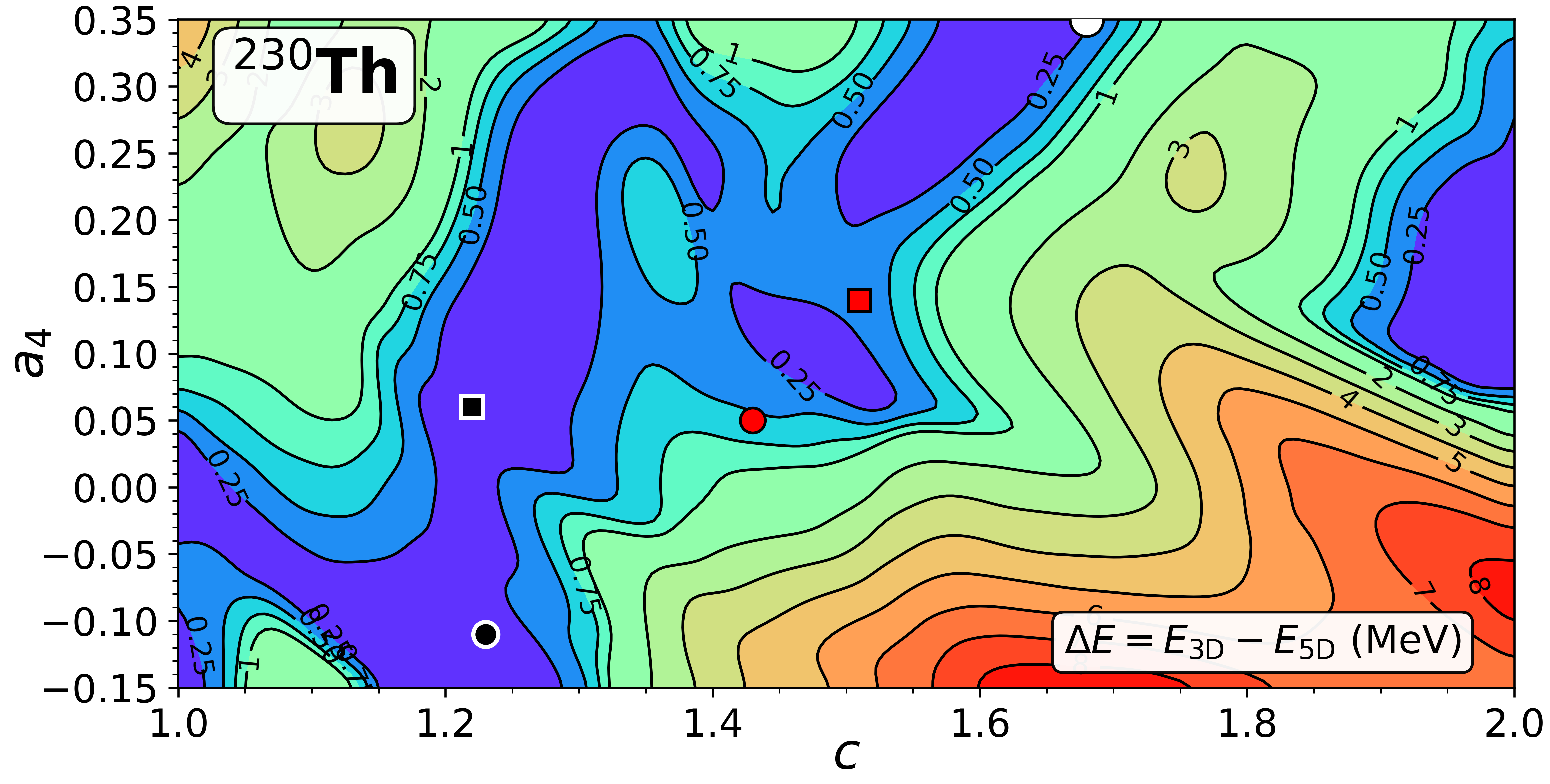}
    \label{fig:230th:diff}
  \end{subfigure}
  \hfill
  \begin{subfigure}[b]{0.49\textwidth}
    \centering
    \includegraphics[width=\textwidth]{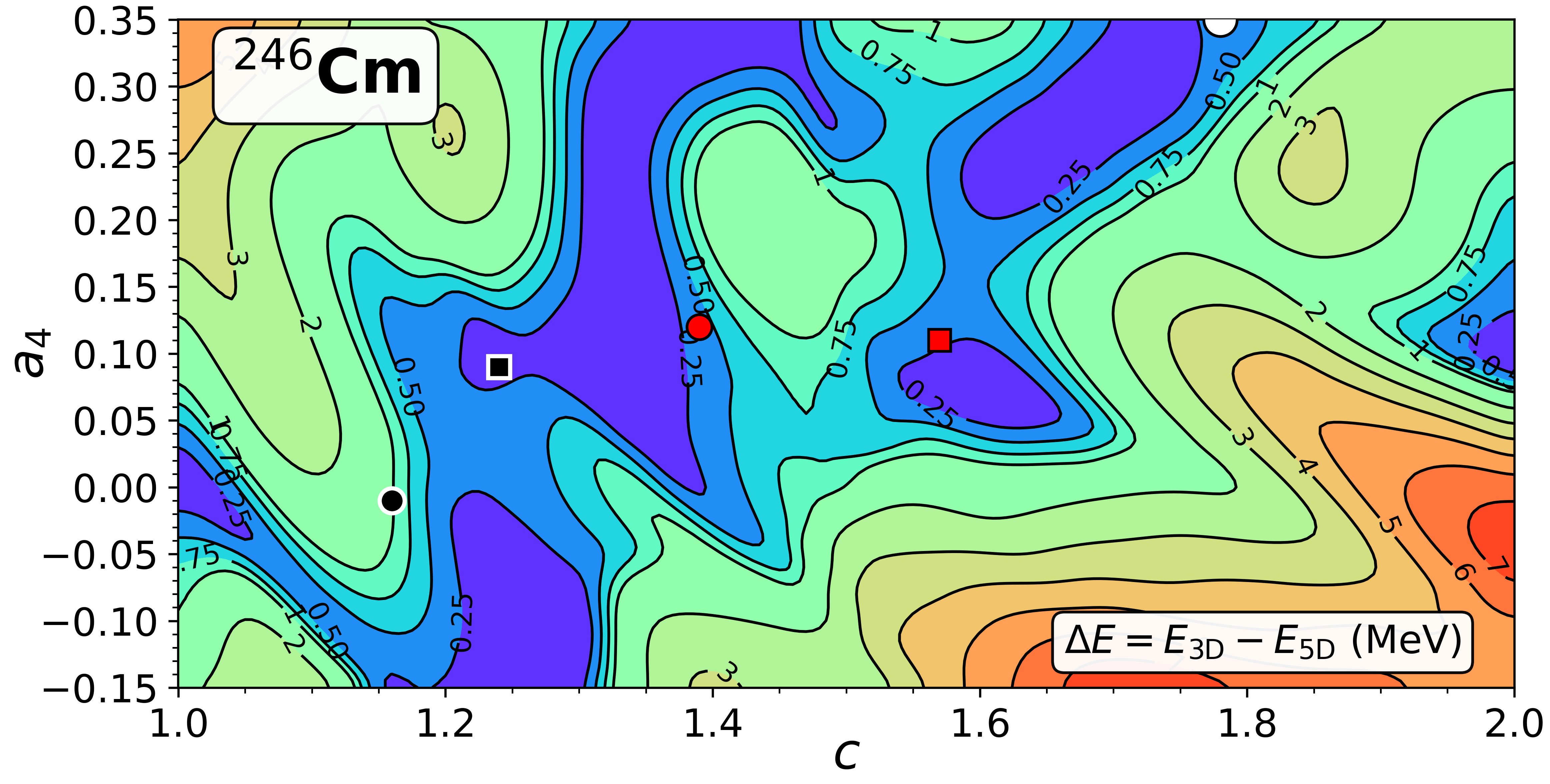}
    \label{fig:246cm:diff}
  \end{subfigure}
  \caption{Map of the energy difference $\Delta E = E_{\text{3D}} - E_{\text{5D}}$ for \ce{^{230}_{90}Th} and \ce{^{246}_{96}Cm}, where $E_{3D}$ and $E_{5D}$ represent the PES calculated in the $(c, a_3, a_4)$ subspace and the full $(c, a_3, a_4, a_5, a_6)$ FoS space, respectively. The color scale (in MeV) shows the local energy difference resulting from including two additional deformation degrees of freedom ($a_5$, $a_6$) beyond the three-dimensional subspace. Circles and squares denote the ground state, the secondary minimum, the fission exit, and the inner and outer saddles on the 5D energy surface as in Fig.~\ref{fig:maps}.}
  \label{fig:diff_maps}
\end{figure*}

A deeper insight into the role of higher-order shape degrees of freedom is provided by the difference maps shown in Fig.~\ref{fig:diff_maps}, which display the quantity $\Delta E = E_{3D} - E_{5D}$ for $^{230}$Th and $^{246}$Cm. These maps quantify the energy gain achieved when the deformation space is extended from the reduced $(c,a_{3},a_{4})$ subspace to the full five-dimensional FoS parameterization that also includes $a_{5}$ and $a_{6}$.

At first glance, one observes a broad blue region covering the ground state, both saddle points, and the fission valley, indicating that in these parts of the PES, the influence of the higher multipoles is relatively small; Therefore, a preliminary exploratory calculation could indeed be performed without them.  However, a more detailed inspection reveals several important effects.

A particularly striking feature appears already at the ground state of $^{246}$Cm. The black circle marking the equilibrium point lies in a region where $\Delta E$ exceeds $1$~MeV, demonstrating that the three-dimensional parameterization fails to reach the correct minimum even near the compact configuration. This is in line with Patyk and Sobiczewski's observation, who showed the importance of $\beta_6^0$ and $\beta_8^0$ deformations in ground-state mass evaluations~\cite{Patyk1991}. Thus, in heavy actinides such as Cm, the higher Fourier components play an essential role not only at large elongations, but also in the vicinity of the ground state, and their omission would lead to a systematic overestimation of the total binding energy by more than one MeV.

An even more surprising behavior is observed above the second saddle point of ${}^{246}\mathrm{Cm}$.  In this physically relevant region, the difference $\Delta E$ continues to grow and reaches approximately $1$~MeV around $c \approx 1.5$ and $a_{4} \approx 0.2$.  This indicates that the additional deformation coordinates strongly stabilize the nucleus already in the early phase of the post-barrier descent. Such a strong and persistent effect of higher-order shape terms is entirely absent in ${}^{230}\mathrm{Th}$, where the same deformation region exhibits only a weak sensitivity to the inclusion of $a_{5}$ and $a_{6}$.  The comparison between the two nuclei therefore reveals a qualitatively different topography of the potential-energy surface in heavier actinides and confirms that a reduced deformation space may, in some cases, be insufficient to accurately capture the landscape beyond the outer barrier.

Overall, the difference maps underscore that higher-order Fourier components are essential for a faithful minimization of the potential energy in a wide range of deformations.  They influence not only the hyperdeformed configurations and the region near scission, but also the ground state area and the entire post-saddle descent.  

\subsection*{Discussion of the third minimum in \texorpdfstring{$^{230}$Th}{230Th}}

The third shallow hyper-deformed minimum in the potential energy of light actinides was predicted more than four decades ago by M\"oller \textit{et al.} \cite{Moller72,Moller72a,Moller73}. The third minima in actinides were experimentally confirmed by measuring rotational level structures at 5 to 6 MeV of excitation energy, with spacings corresponding to moments of inertia consistent with the hyper-deformed calculated minimum \cite{Blons,Blons1,Blons2,Blons3,Blons4}.

The existence of a third hyper-deformed minimum in actinide nuclei has been well established through modeling of measured fission cross sections in neutron-induced reactions, as demonstrated by Sin \textit{et al.} \cite{Sin06,Sin17,Sin21} using the optical model for fission \cite{Sin08,Sin16}. Tables of empirically derived triple-humped fission barrier heights and well depths are available for $^{232}$Pa and $^{233}$Th in \cite{Sin06}, and for light U isotopes in \cite{Sin17}. All these nuclei exhibit a shallow hyper-deformed minimum, with a well depth of less than 1~MeV.

In stark contrast to the above statements, the existence and depth of the third, hyper-deformed minimum in many modern PES calculations for actinide nuclei remain elusive, owing to long-standing and subtle issues in the theoretical description of fission. In our earlier work within the Warsaw macroscopic–microscopic model~\cite{KowalSkalski2012}, which employed a multipole expansion of the nuclear surface up to $\lambda = 8$ (including non-axial terms), the inclusion of the dipole deformation $\beta_{1}^{0}$ effectively eliminated spurious deep hyper-deformed minima reported in previous studies, leaving only very shallow third wells in $^{230,232}$Th, typically not exceeding 200–300~keV in depth.

The present study allows us to investigate how the choice of shape parametrization affects the depth of the third minimum by employing the FoS parametrization while all other model components remain unchanged: the macroscopic–microscopic framework, Woods–Saxon potential, and macroscopic energy prescription remain identical to previous work. As shown in Fig.~\ref{fig:maps}(a), the potential-energy surface of $^{230}$Th reveals a weak but distinct third minimum located at $c \approx 1.6$ and $a_4 \approx 0.2$, with a depth of about 0.5~MeV — only slightly deeper than that found in Refs.~\cite{KowalSkalski2012}. 
The most probable fission path in this parametrization involves rapid increase in neck constriction rather than gradual elongation. Consequently, the path to scission is more vertical in the $(c,a_4)$ plane, indicating that the nucleus tends to develop a thin neck at moderate elongations rather than stretching extensively before rupture. In the case of $^{230}$Th, the third hyperdeformed minimum located around $c \approx 1.6$ and $a_4 \approx 0.2$ forms a shallow pocket with a depth of the order of a few hundred keV, in line with the $\sim 0.5$~MeV estimate discussed above. On the exit side, the third minimum is separated from more compact configurations by a slightly higher saddle; however, the physically relevant fission path in actinides does not proceed toward the symmetric valley but rather follows the direction of increasing mass asymmetry. A descent toward the symmetric valley would require the system to move along configurations with nearly constant neck thickness and increasing elongation. Although this symmetric route is energetically higher, the corresponding barrier is relatively short and steep. By contrast, the asymmetric pathway — developing toward configurations with a rapidly thinning neck and a more pronounced elongation — lies energetically lower but is characterized by a broader and more extended ridge. The interplay between a higher yet narrow symmetric barrier and a lower but wider asymmetric barrier may therefore influence the ultimate choice of the collective trajectory. Thus, one cannot exclude that dynamical effects, not incorporated in the present static analysis, may further modify the preferred direction of motion. In particular, the effective topography of the potential-energy surface may change once the deformation dependence of the centrifugal barrier is taken into account, since its magnitude varies across the multidimensional shape space as the nucleus evolves from compact to more elongated configurations. 
A detailed study of these dynamical corrections and their impact on the accessibility and stability of the third minimum is left for future work.

The comparison with our previous multipole-based and Cassinian-oval analyses~\cite{Jachimowicz2024} thus suggests that the presence and apparent depth of the third minimum are primarily governed by the geometric completeness of the adopted shape parametrization. All three approaches share the same macroscopic–microscopic model, potential, and fitted parameters; hence, the differences arise solely from how the nuclear surface is represented. It should be emphasized that this problem is not merely of an academic nature. Even modest changes in the topography of the outer barrier and the hyperdeformed region can modify the calculated transmission coefficients and statistical fission probabilities, with direct implications for evaluated fission cross sections and class-III resonance predictions. In heavier actinides, such as U and Pu isotopes, the situation is even more pronounced. Within the same FoS formalism, no clear trace of a hyperdeformed third minimum is observed — the potential energy monotonically decreases toward scission after the outer saddle. This finding is particularly intriguing in the case of light U isotopes, where several model calculations of the measured fission cross sections~\cite{Sin06,Sin17} require the presence of a well-developed shallow third minimum to reproduce experimental resonance structures. Our results support the empirical estimates of shallow third minima, but disagree with other studies that have found much deeper, hyper-deformed minima. 

\section*{Summary and Conclusions}

A systematic study of fission barrier heights and static properties of even–even actinide nuclei from Th to Cf has been performed within the Warsaw macroscopic–microscopic model using the five-dimensional Fourier-over-Spheroid (FoS) shape parameterization. The adopted approach, characterized by geometric completeness and high numerical resolution, enables a consistent description of ground states and saddle configurations without requiring any additional parameter adjustments. The large deformation grid of approximately $1.3\times10^{8}$ points for each nucleus ensures full numerical stability and eliminates interpolation-related uncertainties.

The calculated ground-state mass excesses show excellent agreement with experimental data, with a mean deviation of $+0.11$~MeV and a root-mean-square (RMS) error of $0.15$~MeV. The fission barrier heights reproduce empirical evaluations with mean deviations below $1$~MeV. Specifically, the inner barriers are overestimated on average by $+0.46$~MeV (RMS~$1.53$~MeV), while the outer barriers show a smaller mean deviation of $+0.39$~MeV (RMS~$0.64$~MeV). These results confirm the quantitative reliability of the present implementation and demonstrate the robustness of the Warsaw macroscopic–microscopic model when applied to the five-dimensional FoS deformation space.

Special attention has been devoted to the long-debated problem of the third, hyperdeformed minimum. For Th isotopes, a shallow but distinct third well is observed, with a depth of approximately 0.5 MeV. In contrast, this minimum disappears entirely in heavier actinides, such as U and Pu. This finding contrasts with phenomenological model calculations of fission cross sections, which require deeper wells to reproduce fission-resonance data in light U isotopes, and suggests that the existence and depth of this feature depend sensitively on the adopted shape parameterization rather than on the macroscopic–microscopic framework itself. The issue remains open and is of considerable importance for accurately modeling fission dynamics and measured fission cross sections. The systematic survey of the actinide region performed in this work shows that the emerging topography of the five-dimensional PES is broadly consistent with empirical expectations concerning the occurrence of shallow third minima. In particular, for Th isotopes, the calculated structures agree qualitatively with the three-humped fission-barrier pattern suggested by empirical systematics, whereas for U, no such third minimum has been identified. This discrepancy with empirical fission–model reconstructions for U indicates that additional effects (possibly related to dynamical corrections, or the deformation dependence of the centrifugal barrier) may be required to fully understand the experimental systematics in this region. For Pu isotopes, on the other hand, the absence of a third minimum is fully consistent with both empirical estimates and the present macroscopic–microscopic calculations. Further work, combining static PES analysis with multidimensional dynamical modelling, will be necessary to clarify the origin of the difference between Th and U nuclei and to assess the physical relevance of hyperdeformed configurations in medium-heavy actinides.

Overall, the present study demonstrates that the extended Fourier-over-Spheroid parameterization significantly improves the consistency of fission-barrier predictions across isotopic chains and captures the subtle shape effects that govern nuclear stability. Future extensions of this work will focus on several complementary directions. First, the study will be extended to odd-$A$ and odd–odd nuclei. Second, triaxial degrees of freedom, which are known to lower the inner barrier, will be incorporated to improve the description of the first saddle configuration. Third, we plan to extend our analysis by evaluating the fission barrier heights employing two alternative formulations of the macroscopic energy: the Lublin–Strasbourg Drop (LSD) model~\cite{Pomorski2003} and the ISOscalar Liquid Drop Approximation (ISOLDA)~\cite{Pomorski2025}. Both approaches represent distinct phenomenological realizations of the macroscopic component within the macroscopic–microscopic framework. Preliminary estimates presented in Ref.~\cite{Pomorski2025} suggest that these two models are capable of reproducing not only the ground-state binding energies across the entire nuclear chart but also the systematic trends of fission barriers. By incorporating the LSD and ISOLDA prescriptions into the present five-dimensional Fourier-over-Spheroid framework, we will be able to assess the sensitivity of the calculated barrier heights to the chosen macroscopic energy functional and to quantify the model dependence of the predicted barrier systematics. 

Finally, the static potential energy surfaces obtained here will serve as input for multidimensional Langevin and Brownian shape-motion simulations, enabling the calculation of fragment-mass distributions, pre-scission neutron multiplicities, and fission timescales within a unified dynamical framework.

\begin{acknowledgments}
Our research was supported in part by the National Science Centre, Poland, under research project No. 2023/49/B/ST2/01294. M.K. and T.C. were partially co-financed by the COPIGAL Project. 
\end{acknowledgments}

\bibliography{fos_to_beta_barriers_prc}

\end{document}